\magnification=1200
\pageno=1
\centerline {\bf P-BRANE QUANTUM MECHANICAL WAVE EQUATIONS    }
\centerline {\bf    }
\bigskip
\centerline { Carlos Castro}
\centerline { Center for Theoretical Sudies of Physical Systems}
\centerline { Clark Atlanta University, Atlanta, GA.30314}
\centerline { Asian Pacific Center for Theoretical Physics}
\centerline {Seoul, Korea}
\smallskip
\centerline { December 1998}
\bigskip
\bigskip

\centerline { \bf ABSTRACT}
Several quantum mechanical wave equations for $p$-branes are proposed based on the role that the volume-preserving diffeomorphisms group has on the physics of extended objects. The $p$-brane quantum mechanical wave equations determine the quantum dynamics involving the creation/destruction  of $p$-branes in a $D$ 
dimensional spacetime  background  with a given world-volume measure configuration in a given quantum state $\Psi$.

\bigskip

\centerline{\bf I. {INTRODUCTION}}
\bigskip

Quantization of $p$-branes is a notoriously difficult unsolved problem. 
We present here four different quantum mechanical wave equations associated with the quantum dynamics of $p$-branes based on the view that $p$-branes can be seen as gauge theories of volume preserving diffeomorhisms. Three of these wave equations represent the quantum field theory dynamics of $p$-branes living in a $D$ dimensional spacetime with a given world volume measure configuration in a quantum state $\Psi$. The last wave equation is a step towards a finite dimensional formulation of the field theory of $p$-branes.  

We hope to provide a modest step into the solution of the $p$-brane quantization problem. The field theory, represented by the wavefunctional $\Psi$,  that we are advocating here deals with the creation/destruction of $p$-branes histories or $p+1$-world volume measure configurations in a given 
quantum state $\Psi$. This will permit us to write down the wave functional equations. 

\bigskip
\centerline{\bf II. p-Brane Wave Equations}
\bigskip

\centerline{\bf 2.1 Wave Equations based on Antisymmetric Tensor Field Theories}
\bigskip
A local antisymmetric tensor gauge field theory reformulation of extended objects was given by [1,2] which later paved the way for the author to show that $p$-branes could be seen as composite antisymmetric tensor gauge field theories of the volume-preserving diffeomorphisms group [3,4] associated with the target  space of the primitive scalar field constituents . The action [1,2] was defined  :

$$S=-g^2 \int d^D x \sqrt { - {1\over (p+1)!}
(W_{\nu_1 \nu_2....\nu_{p+1}})^2 } +      
{1\over (p+1)!}W^{\nu_1 \nu_2....\nu_{p+1}} \partial_{\nu_1} B_{ \nu_2....\nu_{p+1}}. \eqno (2.1)$$
$ B_{ \nu_2....\nu_{p+1}}$ was a Lagrange multipier enforcing the transversality condition on the antisymmetric current $ W^{\nu_1 \nu_2....\nu_{p+1}} $. 
The field equations associated with the $W$ and $B$ fields admit $p$-brane solutions when the antisymmetric tensor current $W$ evaluated on any spacetime point $x^\mu$, is related to the $p+1$-vector tangent to the worlvolume of the $p$-brane  :

$$ W_{\nu_1 \nu_2....\nu_{p+1}}(x) =\kappa \int d^{p+1}\sigma 
\delta^D (x -X(\sigma )) {\vec X}_{\nu_1 \nu_2....\nu_{p+1}} (\sigma). \eqno (2.2)$$
with the $p+1$-vector tangent to the $p$ brane's worldvolume  :

$$ {\vec X}_{\nu_1 \nu_2....\nu_{p+1}} (\sigma)=\epsilon^{\sigma^1 \sigma^2....\sigma^{p+1}}\partial_{\sigma^1}X_{\nu_1}......\partial_{\sigma^{p+1}}X_{\nu_{p+1}}. \eqno (2.3)$$
Plugging   the solution (2.2, 2.3) into the original action (2.1) yields the $p$-brane Dirac-Nambu-Goto action after using the property of the delta function integration :

$$S=-(\kappa g^2) \int d^{p+1}\sigma \sqrt { - {1\over (p+1)!}
({\vec X}_{\nu_1 \nu_2....\nu_{p+1}})^2 }. ~~~T_p=\kappa g^2. \eqno (2.4)$$
the $p$-brane tension is expressed in terms of the coupling constant $g$ with the addition of an extra parameter $\kappa$ to accomodate dimensions : $ T_p=\kappa g^2$. The solution (2.2) is defined up to a scaling parameter given by $\kappa$.  
The $p$-brane extension on the Hamilton-Jacobi equation is [1,2] :

$$- {1\over (p+1)!} ({\Pi}_{\nu_1 \nu_2....\nu_{p+1}})^2 +T^2_p = 
- {1\over (p+1)!} [ {\partial (S^1(x), S^2(x),...,S^{p+1})\over      
\partial (x^{\nu^1}, x^{ \nu^2},....x^{\nu^{p+1}} ) } ]^2 +T^2_p =0. \eqno (2.5)$$
where $S^1(x), S^2(x),...S^{p+1}(x)$ are the Clebsch Potentials [1,2]. The quantity ${\Pi}_{\nu_1 \nu_2....\nu_{p+1}}$, when extended from the $p$-brane's world tube to all of spacetime, is the so-called slope or geodesic field of the $p$-brane.  The pullback from spacetime to the $p$-brane's world tube of the slope/geodesic field is the $p$-brane's volume-momentum, 
$\Pi^{a^1}_{\nu_1} (\sigma^a)$    ,
conjugate  to the $p$-brane's configuration coordinate $X^\nu (\sigma^a), ~a=1,2,...p+1$ :

$${\Pi}_{\nu_1 \nu_2....\nu_{p+1}} \epsilon^{a^1 a^2....xa^{p+1} }
\partial_{a^2}X^{\nu_2}(\sigma^a) \partial_{a^3}X^{\nu_3}(\sigma^a)......... 
\partial_{a^{p+1}} X^{\nu_{p+1}} (\sigma^a) \equiv \Pi^{a^1}_{\nu_1} (\sigma^a). \eqno (2.6)$$
 
The reformulation of $p$-branes in terms of antisymmetric tensor fields was based on the old vortex/hydrodynamics ideas of Nielsen and Olesen.    
$p$-branes are reinterpreted as higher dimensional vortices. The vortex velocity (inside the fluid) at any given point of the vortex matches the fluid's velocty at that given point. The vortex velocity plays the role of the $p+1$-vector tangent to the $p$-brane's world tube whereas the fluid's velocity plays the role of the slope/geodesic field, ${\Pi}_{\nu_1 \nu_2....\nu_{p+1}}$ defined outside the $p$-brane. When the slope/geodesic field is restricted to have support on the $p$-brane's world tube then it matches the $p+1$-vector tangent to the $p$-brane.         

We propose our first quantum mechanical wave functional equation for the $p$-brane based on the Hamilon-Jacobi equation to be of the following  form  :

$$ { 1 \over (p+1)!} { \delta^2 \Psi \over  \delta {\vec X}_{\nu_1 \nu_2....\nu_{p+1}} \delta {\vec X}^{\nu_1 \nu_2....\nu_{p+1}} } +
{T^2_p\over  m^{2(p+1)}         }\Psi =0. \eqno (2.7)$$
which is just the constraint implementation of the Hamilton-Jacobi on-shell condition (2.5)  on $\Psi$. We set our units to be such $\hbar =1$ and $m$ is a suitable parameter of mass needed to match dimensions . This is $not$ an artificial introduction of an extra parameter, $m$, but it has a well defined physical meaning in the Eguch-Schild quantization scheme of the string [5]. The usual Nambu-Goto-Polyakov quantization is based on keeping the string tension constant while taking the average over all worldsheet areas. The string tension is related to the Planck scale/Regge slope $\alpha '$  : 

$$T_2=m^2={1\over 2\pi \alpha '}. \eqno (2.8)$$ 
Whereas in the Eguchi-Schild quantization one starts from an action given by the squared of the string's worldsheet area ( instead of the area) and invariant under area-preserving diffs :

$$S = m^2 \int d^2\sigma \{ X^\mu, X^\nu \}_{PB}\{ X_\mu, X_\nu \}_{PB}         \eqno (2.9)$$
the Poisson brackets are taken w.r.t the $\sigma^1,\sigma^2$ variables. 
One quantizes by keeping fixed the world sheet area of the string histories 
and taking the average over all the $variable$ string tension values , $T$; i.e the energy per unit length of the string is not kept constant. The Nambu-Goto quantization imposes the condition : $m^2=T$ which is not the case in the Eguchi-Schild quantization scheme. 

Therefore, $m^{2(p+1)}$ is not necessary equal to $T^2_p$.  
For these reasons we shall set the ratio $(T_p/m^{p+1})^2 $ to be equal to 
$\lambda^2$; i.e the ``eigenvalue'' of the functional derivative operator acting on $\Psi$.   
The quantity : 

$$\Psi=\Psi [ {\vec X}^{\nu_1 \nu_2....\nu_{p+1}} (\sigma^1,....,\sigma^{p+1})]. \eqno (2.10)$$  
is the wave $functional$ associated to a $p$-brane with a given $p+1$-vector tangent to the $p$-brane, ${\vec X}^{\nu_1 \nu_2....\nu_{p+1}} (\sigma^a)$ (  
with a given $p+1$ world volume $p$-brane measure)  in the quantum state $\Psi$. We must emphasize that this wavefunctional must not be interpreted as a probability ampitude. We are dealing with a truly second-quantized field theory of $p$-branes  and $not$ a first quantized one.   
$\Psi$ truly represents a field that creates/destroys a $p$-brane with a given $p+1$ world volume measure configuration ( $p$-brane's history) embedded in  a $D$-dimensional target spacetime background in the quantum state $\Psi$. $p$-branes are seen here as theories related to gauge theories of the volume-preserving diffs group. The latter are not ordinary non-abelian YM gauge theores that we are familiar with but composite antisymmetic tensor field theories [3]. 

In [3] we have shown that $p$-branes can been seen as extended solutions to composite antisymmetric tensor gauge field theores of the volume preserving diffs group. The actions were of the YM types which precisely bear the same resemblance as the Schild-action associated with the group $SU(\infty)$ of area-preserving diffs for spacetime independent gauge field configurations ( vacuum). The actions [3] were :

$${1\over g^2_p}[{F}_{\nu_1 \nu_2....\nu_{p+1}} (\phi^a (x)) ]^2 . \eqno (2.11)$$
with the $composite$ field strength defined in [4] : 
$${F}_{\nu_1 \nu_2....\nu_{p+1}} (\phi (x)) = \epsilon _{a_1 a_2......a_{p+1}}
\partial_{\nu_1} \phi^{a_1}(x) \partial_{\nu_2} \phi^{a_2}(x)....
\partial_{\nu_{p+1}}\phi^{a_{p+1}}(x). \eqno (2.12)$$
the fields $\phi^a (x)$ are a collection of $a=1,2,....p+1$ primitive scalars
living in a spacetime of dimensionality  $D=p+1+p'+1$ and taking values in a target space 
of dimensionality $p+1$. The $dual$ primitive scalars ${\tilde \phi}^b(x)$  
live in spacetime and take values in a target space of dimensionality $p'+1$.
Notice that this is the $dual$ picture to what we are familiar with  in the description of $p$-branes. Here we have maps from spacetime into manifolds of $less$ dimension : an inmersion. We have shown  that  every strongly coupled composite antisymmetric tensor field theory of $p+1$-rank defined in $D$ spacetime has a weakly coupled $p'$ brane solution and vice versa. Also we have $T$ duality 
with a large/small volume duality in the space of solutions . $S$ and $T$ dualities were interrelated to each other. 
The relevance of these composite theories is that we have a formulation of $p$-branes where $S, T$ dualities are already 
built in from the very start : at the Lagrangian level, there is no need to conjecture them. 
 
There was another sort of duality [3] that could be inferred under the exchange: 

$$\partial_{\sigma^a} X^{\mu_1} \leftrightarrow \partial_{\mu_1} \phi^a (x). \eqno (2.13)$$  
this duality (2.13) exchanges target manifold  for base manifold. The composite field strength is then replaced by :
$$[{F}_{\nu_1 \nu_2....\nu_{p+1}} (\phi^a (x)) ]^2 
\leftrightarrow [\epsilon^{\sigma^1 \sigma^2....\sigma^{p+1}}\partial_{\sigma^1}X^{\nu_1}......\partial_{\sigma^{p+1}}X^{\nu_{p+1}}]^2. \eqno (2.14)$$
ths last equation (2.14) has the similar  form as the Skyrmion-based Dolan-Tchakrian conformaly invariant Lagrangians [6]  for the $p$-brane in the case that $p+1=2n$. Similar actions can also be constructed for $p+1=2n+1$ but in such case conformal invariance was lost. A spinning membrane action based on the Dolan-Tchrakian Lagrangians  was constructed in [7]. 
Dolan-Tchrakian have shown that upon the algebraic elimination of the auxiliary world volume metrics $g^{ab}$, via their equations of motion,  one recovers the Nambu-Goto actions. The antisymmetrization w.r.t the indices of Dolan-Tchrakian's  action was done as follows :

$$g^{\sigma^1 \xi^1 }g^{\sigma^2\xi^2 }....g^{\sigma^{p+1}\xi^{p+1} }
\partial_{[\sigma^1}X^{\mu_1}......\partial_{\sigma^{p+1}]}X^{\mu_{p+1}} \partial_{[\xi^1}X_{\mu_1}......\partial_{\xi^{p+1}]}X_{\mu_{p+1}}.\eqno (2.15)$$
the antisymmetrizations of indices $[\sigma^1,\sigma^2....\sigma^{p+1}]$  and 
$[\xi^1,\xi^2....\xi^{p+1}]$
will render such actions in the similar form  of eq-(2.14). One can either antisymmetrize the sums of products of world volume metric indices or the derivatives acting on the coordinates $X^\mu$.  The latter Skyrmion-based actions are of the $p+1$-volume squared form ( like the Schild action) . 

\bigskip
  
\centerline{\bf 2.2. The Loop Space Wavefunctional of a $p$-brane} 
\bigskip

A noncovariant Schroedinger-like loop wave equation for the bosonic closed string based on the Eguchi $areal$ time  quantization scheme 
of the Schild action was given by [5] :
$$H\Psi [\sigma^{\mu\nu}, A] =-i {\partial \Psi \over \partial A}. ~~~\hbar =1.~~~
\Psi =\Psi e^{-i {\cal E} A}. $$
$$- {1\over l_C} \int_0^1 ds \sqrt {x'(s)^2 }{1\over 4m^2} ({\delta^2 \over 
\delta \sigma^{\mu\nu} (C) \delta \sigma_{\mu\nu} (C) } \Psi [\sigma^{\mu\nu}(C)])=
{\cal E}\Psi [\sigma^{\mu\nu}(C)]. \eqno (2.16)$$     
the string wave functional in this nonrelativistic Schroedinger-like loop wave equation represents the quantum amplitude to find a given closed loop $C$, with spatial area-elements or $holographic$ shadows $\sigma^{\mu\nu}(C)$  onto a spacetime background, as the only boundary of a two-surface of internal area $A$ in a given quantum state $\Psi$. ${\cal E}$ is the energy per unit length or the string tension and $m$ ( discussed earlier)  is the  mass parameter related to the string Regge slope which is not necessary  equal to the variable string tension : ${\cal E}$. 
$l_C$ is he reparametrization invariant length of the loop $C$.

The $covariant$ Loop wave equation must treat the $areal$ time $A$ on the same footing as the $spatial$ area or holographic components $\sigma^{\mu\nu}(C)$. This led the author [8] to write a Klein-Gordon like Master field equation in a generalized {\bf C}-space where 
point, loops and surface histories were $all$ treated on a single footing and where the extension of the Special Relativiy principle transformed each object 
into one another; i.e the point, loop and surface histories are  reshuffled into each other under the generalized Lorentz transformations which in {\bf C}-space amount to area-preserving antisymmetric matrix-coordinates transformations. 

This is consistent with the ideas of Poly-Dimensional-Relativity of [9] that purports to treat all dimensions and signatures of spacetime into one single framework of multidimensional or Clifford-algebra valued equations . Relativistic transformations in Poly-Dimensional Relativity  transform a {\bf 2}-vector into a {\bf 3}-vector or into a {\bf 5}-vector, for example. This occurs also under paralell transport of a {\bf p}-vector along a closed curve : the rank of the {\bf p}-vector can change. Dimensions/signatures are on the eye of the beholder. It has been argued that duality in 
Electromagnetism is linked to signature change rotations [9].     

The covariant loop wave functional equation for histories taken place in $D=4$ required the use of an $eleven$ dimensional  {\bf C}-space of $8\times 8$ antisymmetric matrices, $X^{MN}$ ( whose entries have  area-dimensions ) encoding the point,loop and surface histories :

$$  {\delta^2 \over 
\delta X^{MN} \delta X_{MN}  } \Psi [X^{MN}]+T^2 \Psi [X{MN}] =0. \eqno (2.17)$$
In general, for $p+1$ branes one must use an antisymmetric tensor of rank 
$p+1$ instead of the antisymmetric matrix $X^{MN}$ and replace the string tension squared terms with the $p$-brane tension squared. The result of (2.17) also includes $tensionless$ or null strings. The analog of a ``photon'' in {\bf C}-space are the null strings/null $p$-branes. 

Schroedinger-like loop wave equations for $p$ branes were also discussed in 
[8]. $p$-branes of topologies $S^p$ had for higher-dimensional loop wave equations exactly the same form as (2.16) where now one must add indices to all the quantities and replace the variable string tension, ${\cal E}$,  for the $p$-brane one : Energy per unit volume. The holographic coordinates $\sigma^{\mu\nu} (C)$ become now :
$\sigma^{\mu_1....\mu_{p+1}} (S^p)$ expressing the proyection of the volume components enclosed by the $p$-brane ( of topology $S^p$)  onto spacetime.   
Pavsic [10] has constructed wave equations for ``wiggled'' $p$ branes based on a Fock-Schwinger 
proper time unconstrained formalism at the expense of introducing auxiliary fields ( the wiggles). His construction also relied on the loop space approach method.  

\bigskip
\centerline{\bf 2.3 Supersymmetric Quantum Mechanical Wave Equations for $p$-Branes}
\bigskip

A Dirac-square root like procedure for the Hamilton-Jacobi inspired equation 
(2.7) yields :

$$i \Gamma^{\mu_1 \mu_2......\mu_{p+1}} {\delta \Psi \over \delta 
{\vec X}^{\mu_1 \mu_2......\mu_{p+1}} } +{T_p\over m^{p+1}} \Psi =0. \eqno (2.18) $$
where we have inserted again a length/mass scale parameter $m$ to adjust 
dimensions. The anticommutator of the antisymmetric 
$\Gamma^{\mu_1 \mu_2......\mu_{p+1}} $ matrices is proportional to the unit matrix $I$ : 

$$\{\Gamma^{\mu_1 \mu_2......\mu_{p+1}}, \Gamma^{\nu_1 \nu_2......\nu_{p+1}}\}={\cal G}^{ ([\mu_1 \mu_2......\mu_{p+1}][\nu_1 \nu_2......\nu_{p+1}] )} I. 
\eqno (2.19)$$
with 
$${\cal G}^{( [\mu_1\mu_2......\mu_{p+1}][\nu_1 \nu_2......\nu_{p+1}]) } =
g^{\mu_1 \nu_1}g^{\mu_2 \nu_2}......g^{\mu_{p+1}\nu_{p+1}} + permutations. \eqno (2.20)$$
the permutations ensure that the generalized metric is antisymmetric under any exchange of $\mu_1, \mu_2.....\mu_{p+1}$ and $\nu_1, \nu_2.....\nu_{p+1}$ indices, respectively,   and is symmetric under the exchange of the collective group indices $\mu \leftrightarrow \nu$. Notice that the sum of antisymmetrized  products of 
ordinary metrics appears also in the Dolan-Tchrakian action (2.14,2.15).    

The matrices $\Gamma^{\mu_1 \mu_2......\mu_{p+1}}$ can be written as linear combinations of sums of antisymmetrized products of ordinary Dirac-Clifford algeba matrices :

$$\Gamma^{\mu_1 \mu_2......\mu_{p+1}} =\sum_{i=1}^{2^D} C^{\mu_1 \mu_2......\mu_{p+1}} _{(i)} \Gamma^{(i)}. \eqno (2.21)$$
where a $2^D$ dimensional Clifford algebra basis is spanned by the antisymmetrized products of the ordinary $2^{D/2} \times 2^{D/2}$ $\gamma^\mu$  matrices :   

$$ \Gamma^{(i)} = \{ I, \gamma^{\mu_1}, \gamma^{[\mu_1}\gamma^{\mu_2 ]},....
  \gamma^{[\mu_1}\gamma^{\mu_2 }\gamma^{\mu_3}.....\gamma^{\mu_D ]} \}. 
\eqno (2.22)$$
Eqs-(2.19, 2.20, 2.21, 2.22) will in principle determine  the algebraic equations to solve for the  $\Gamma^{\mu_1 \mu_2......\mu_{p+1}}$ matrices based on the fact that 
each metric component in the r.h.s of eq-(2.20) can be replaced by the anticommutator 
$\{\gamma^\mu, \gamma^\nu \} =g^{\mu \nu} I$.

\bigskip
\centerline {\bf 2.4 The De Donder-Weyl- Kanatchikov-Navarro Approach}
\bigskip

A finite dimensional formulation of  QFT based on the De Donder-Weyl Hamiltonian formalism of the 1930's has been undertaken in the 1990's by Kanatchikov and Navarro [11] where , among other things, the 
quantization rules and equations of motion are covariant. The approach [11] is  based on Lagrangian systems of the general type $L=L(\phi^a, \partial_\mu \phi^a, x^\mu)$ which may include spacetime sources for the fields $\phi^a$.    
The generalized covariant Legendre transform is $H=\pi^\mu_a \partial_\mu \phi^a -L$ associated with the $polymomentum$ field variables 
$$\pi^\mu_a ={\partial L\over  \partial (\partial_\mu \phi^a)} . \eqno (2.23)$$ . 
The  Kanatchikov-Navarro  construction replaces the Schroedinger equation of ordinary Quantum Mehanics with one of the Dirac-like form :

$$i\gamma^\mu \partial_\mu \Psi = {\hat H} \Psi. \eqno (2.24)$$
The units are $\hbar =1$ and , again, a suitable mass/length parameter must be inroduced in the l.h.s to match dimensions. The field operators have  the following correspondence principle with the classical field observables : 

$${\hat \phi}^a \rightarrow \phi^a.~~~{\hat \pi}^\mu_a \rightarrow -i\gamma^\mu {\partial \over \partial \phi^a}. \eqno (2.25) $$

In the study of $p$-branes we firstly  must be careful with the proper use of indices. We see the $p$-brane as a $p+1$-dimensional classical field theory associated with a set of $D$ scalar fields $X^\mu (\sigma^a) $.   
The Kanatchikov-Navarro covariant finite dimensional approach to the quantum field theory of $p$-branes yields for wave equation :

$$\kappa \hbar \gamma^a \partial_a \Psi = {\partial^2 \Psi \over \partial 
{\vec X}^{\mu_1.....\mu_{p+1}} \partial {\vec X}_{\mu_1.....\mu_{p+1}}. } \eqno (2.26)$$ where now we have ordinary derivatives instead of functional derivatives acting on the wave function $\Psi ({\vec X}^{\mu_1.....\mu_{p+1}}; \sigma^1,....\sigma^{p+1} ) $ representing the probability amplitude to find a $p$-brane in the quantum state $\Psi$ with 
a $p+1$ world volume measure equal to the $numerical$ value of           
${\vec X}^{\mu_1.....\mu_{p+1}}$ at each world volume point $P$ with coordinates ( seen by  an observer living on the world volume of the $p$-brane)  given by $\sigma^1,....\sigma^{p+1}$. One must reiterate that $\Psi$ is $not$ a functional of all the possible $p$-brane's world volume measure field $configurations$  but, instead,  one has a function of the $numbers$ that the measure takes at each given point $P=(\sigma^1,....\sigma^{p+1}) $.       

The Kanatchikov-Navarro approach to the $p$-brane quantum mechanical wave equations based on the De Donder-Weyl Hamiltonian formalism is a sort of a finite dimensional slice of the ( infinite dimensional )  quantum field theory of $p$-branes based on functional differential wave equations discussed above. This finite dimensional approach is certainly worth pursuing because it may yield a suitable regularization mechanism to extract finite-valued physically  relevant information from the infinities which plague odinary QFT. Especially when one is dealing with such complicated theories as the QFT of $p$-branes.

\bigskip
\centerline{ \bf Acknowledgements .}
\smallskip
We thank Igor V. Kanatchikov for many useful conversations  in Trieste, Italy about his work on the De Donder-Weyl Hamiltonian formalism approach to field theories. We extend our gratitude to Profs C.Y Lee and Y. M Cho for the invitation to the Asian Pacific Center for Theoretical Physics in Seoul, Korea where this work was completed. 
\smallskip

\centerline {\bf Referenes} 

1. A. Aurilia, A. Smailagic and E. Spallucci : Phys. Rev. {\bf D 47} (1993) 

2536.

A. Aurilia , E. Spallucci : Class. Quant. Grav. {\bf 10} (1993) 1217. 

2. H. Kastrup : Phys. Reports {\bf 101} (1983) 1.

3. C. Castro : Int. Jour. Mod. Phys. {\bf A 13} (8) (1998) 1263. 

4. E. Guendelmann, E. Nissimov and S. Pacheva : `` Volume preserving Diffs 

versus Local gauge Symmetries `` hep-th/9505128.

5. A. Schild : Phys. Rev {\bf D 16} (1977) 1722.

T. Eguchi : Phys. Rev. Lett {\bf 44} (3) (1980) 126.

S. Ansoldi, A. Aurilia and E. Spallucci : `` Loop Quantum Mechanics and 

the Fractal Structure of Spacetime `` to appear in the Jour. of Chaos, Solitons

and Fractals {\bf 10} (1) (1999) special issue on Superstrings, M, F, S ..Theory.

C. Castro, M. S. El Naschie, eds.

6. B. Dolan , D. Tchrakian : Phys. Letters { \bf B 198} (1987) 447.

7. C. Castro : `` The Spinning Membrane and Skyrmions Revisited `` 

hep-th/9707023;  with the JHEP.

8. C. Castro : `` The Search for the Origins of M Theory....'' hep-th/9809102.

9. W. Pezzaglia : `` Polydimensional Relativity, a Classical Generalization 

of the Automorphism Invariance Principle'' in Clifford Algebras and 

Applications in Mathematical Physics `` eds. V. Dietritch. K. Habetha and 

G. Jank. Kluwer Academic Publishers 1997. page 305. gr-qc/9608052.

W. Pezzaglia : `` Should Metric Signature Matter in Clifford Algebra

Formulations of Physical Theories `` gr-qc/9704043.   

10. M. Pavsic : `` The Dirac Nambu Goto $p$-branes as Particular Solutions to a 

General Unconstrained Theory `` Ljubljana IJS-TP/96-10 preprint.

11. I. V. Kanatchikov : `` The De Donder-Weyl Theory and a Hyercomplex 

extension of Quantum Mechanics to Field Theory `` to appear in Rep. Math. Phys 

{\bf 42} (1998).

 I. V. Kanatchikov :  `` From the  De Donder-Weyl Hamiltonian Formalism to 

Quantization of Gravity ``

to appear in the Procc. Int Seminar in Mathematics and Cosmology, Potsdam 1998.

eds. M. Rainer and H. Schmidt. World Scientific 1998.

M. Navarro : `` Towards a Finite Dimensional Formulation of QFT `` 

quant-ph/9805010.

\bye